\documentstyle[prd,aps,floats,epsfig,graphicx]{revtex}

\def\be{\begin{equation}}
\def\ee{\end{equation}}
\def\bea{\begin{eqnarray}}
\def\eea{\end{eqnarray}}

\def\k{\kappa}
\def\l{\lambda}
\def\p{\phi}

\def\bra#1{\left\langle #1\right|}
\def\ket#1{\left| #1\right\rangle}

\begin{document}
\preprint{BROWN-HET-1245, hep-th/0103019}
\draft 

%
%
 
%
\renewcommand{\topfraction}{0.99}
\renewcommand{\bottomfraction}{0.99}
\twocolumn[\hsize\textwidth\columnwidth\hsize\csname 
@twocolumnfalse\endcsname

\title
{\Large Universe Generation from Black Hole Interiors}
\author{Damien A. Easson and Robert H. Brandenberger}
\address{~\\Department of Physics, Brown University,  
Providence, RI 02912, USA. \\ Email: easson@het.brown.edu, rhb@het.brown.edu }
\date{\today} 
\maketitle
\begin{abstract} 
We point out that scenarios in which the universe is born from 
the interior of a black hole may not posses many of the problems of the 
Standard Big-Bang (SBB) model.  In particular we demonstrate that 
the horizon problem, flatness, and the structure formation problem might be solved naturally, not necessarily requiring
a long period of cosmological inflation.  The black hole information loss problem is also discussed.  Our conclusions are completely independent of
the details of general models.
\end{abstract}

\pacs{PACS numbers: 98.80Cq      \hspace{4.7cm} BROWN-HET-1245, hep-th/0103019}]

\vskip 0.4cm
\section{Introduction}

In this note we consider the properties of a daughter universe spawned from the 
interior of a black
hole resting in a mother universe.  We show that such a daughter universe will take on
certain properties of the mother universe. Possible solutions to the 
horizon, flatness, structure
formation and black hole information loss problems are suggested.  
Many cosmological scenarios in which our universe emerges from the interior of a black hole have been proposed
(see e.g. \cite{FMM} - \cite{DKH}).  Here we provide motivation for studying such scenarios by briefly reviewing some of the relevant work.  

In \cite{FMM} and \cite{Morgan}, it was postulated without any
constructive realization (but based on the limiting curvature conjecture) 
that in the high Weyl curvature region near the
Schwarzschild singularity a matching between the Schwarzschild and the
de Sitter metrics as depicted in Figure 1 takes place. A constructive
implementation of the limiting curvature hypothesis was discovered in \cite{MB}.
In this construction, the gravitational action involves higher 
derivative terms parameterized by means of a 
non-dynamical scalar field $\phi$ with a potential $V(\phi)$.  The higher derivative terms are found using the ``Limiting Curvature Construction'', demanding that at high curvatures the metric become locally de Sitter \cite{MB}.
Based on this principle, it was possible to construct a nonsingular
black hole in $1+1$ dimensions \cite{TMB}. This scenario closely resembles the one we have in mind in this paper.\footnote{We 
are currently extending the construction of \cite{MB} to discuss the four-dimensional Schwarzschild black hole.}

In the scenario of \cite{TV}, a study of non-critical string cosmology
in two spatial dimensions reveals a non-singular black hole.  An observer
crossing the black hole's horizon enters into a new universe. A related
phenomenon occurs in
the Pre-Big-Bang (PBB)\cite{PBB} model which is based on the low energy effective action of string theory.  In this scenario the universe starts out in the string perturbative vacuum state,
in a sea of dilaton and gravitational waves.  Quantum fluctuations lead to the formation of black holes in the Einstein frame which correspond to dilaton driven inflationary phases in the string frame.  At late times, the universe resulting from this construction evolves in accordance with the Standard Big-Bang model.

In \cite{Lowe} the authors analyze cosmology within the framework of
the AdS/CFT correspondence.  They consider the motion of a charged domain
wall that separates an external Reissner-Nordstrom region of spacetime from
an interior de Sitter region.  In this context, solutions exist such that a black hole forms with an interior resembling a toy cosmological model.

Finally, in \cite{DKH} a model is constructed where a black hole forms within the scope of a classical
field theory with false vacuum interior and
true vacuum exterior.  Although the metric on the outside of the black hole is 
Schwarzschild, the interior metric is de Sitter.  The action is the Einstein-Hilbert 
action modified by the inclusion of a scalar field and a potential for that scalar field.

Similar ideas were also put forward by Smolin \cite{Smolin}.
He postulates that the universe may have been born out
of the interior of a black hole which in turn is embedded in a parent universe.
Characteristics of
the parent universe are passed on to the daughter universe.  This is a 
cosmological analogy to Darwin's ``survival of the fittest" theory of
natural selection because only universes with conditions suitable to 
form black holes are capable of reproducing.  Although Smolin provides no quantitative realization
of his model, the arguments of \cite{Smolin} are nonetheless intriguing
and address many of the problems of the SBB.

Due to the extensive work in this area it is clear that 
the idea that our universe might be generated from the inside of a black hole
is of great interest, and that several scenarios have been formulated which produce this
exact result.  In this paper we will discuss a number of features that all such 
scenarios have in common.  Note that the conclusions described in this
paper are \it independent \rm of the specific models.  

The basic set up we have
in mind is represented in the Penrose diagrams of Fig. (\ref{pen}).  As is evident from
this diagram, when an observer crosses over the event horizon,  $r$ and $t$
switch roles (the Schwarzschild metric components $g_{tt}$ and $g_{rr}$
change sign).  In (a), the Schwarzschild black hole has a 
singularity at $r=0$.  In (b), the singularity
is replaced by an initial time surface of the de Sitter universe.  In other 
words, the location $r=0$ from the viewpoint of an observer in the mother universe
becomes the initial time surface $t=0$ for an observer in the baby de Sitter universe.   
Hence all of the matter that will fall into the black hole, eventually reaching 
the same place ($r=0$) but at different times
will enter into the baby universe at the same time ($t=0$) but in different places.   

\vskip 0.4cm
\section{Solving Problems of SBB Cosmology}
 
Many of the problems of the SBB \footnote{For a recent review on the problems of
the SBB and inflationary models see \cite{Robert}.} can be solved if the universe is created inside
a black hole which is resting in a parent universe.  In this section we will
suggest how the horizon, flatness and structure formation problems might be solved without requiring a long period of inflation.  It is interesting to note 
however, that all of the scenarios in \cite{FMM} - \cite{DKH} do involve a de Sitter bounce (which is presumably of too short duration to solve the
problems of standard cosmology).

One of the major problems of the SBB model which is solved by inflation is the
horizon problem.  The horizon problem is essentially a problem of causality.
The region over which the CMB is observed to be homogeneous to better than
one part in $10^4$ is much larger than the comoving forward
light cone of an observer at the time of recombination, which is the maximal distance over which
micro-physical forces could have caused the homogeneity.

Inflationary models solve this problem.  During inflation the forward light cone
expands exponentially.  If inflation lasts for a sufficient number of e-foldings,
the forward light cone will be larger than the past light cone at the time of last scattering.

Inflation is not without its own problems.  In particular, inflation is
not able to address the fluctuation problem, super-Planck-scale physics problem, initial
singularity problem and the cosmological constant problem as described in \cite{Robert}.
It is therefore desirable to search for possible alternative solutions to the
problems of the SBB model.
\vspace{0.4cm}
\subsection{The Horizon Problem}

\begin{figure}[htbp]
\includegraphics[angle=0,width=3.5in]{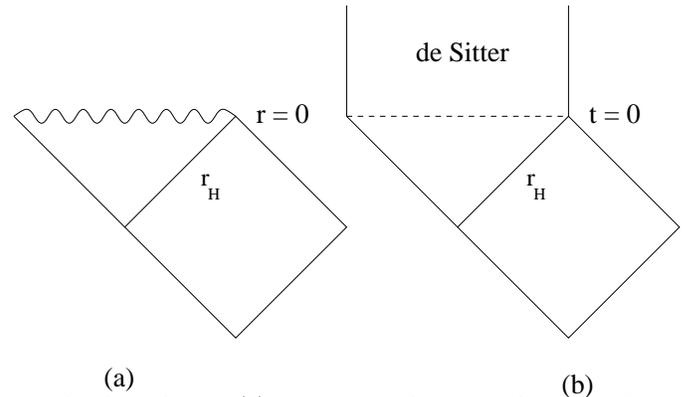}
\caption{Figure (a) shows a Penrose diagram of a Schwarzschild
black hole with event horizon $r_H$ and singularity at $r = 0$.  Figure
(b) shows a nonsingular black hole with de Sitter interior.}
\label{pen}
\end{figure}

Now we will describe how the causality aspect of the horizon problem is solved naturally within the
context of models which are of the type depicted in Fig.(\ref{pen}). First of 
all, note that the forward light cone of a point on the horizon at $t = 0$
(where $t$ is the Schwarzschild time coordinate) encompasses all points along 
the radial direction in the de Sitter universe. All that needs to be shown
therefore is that causal contact is possible along the angular directions.   
Thus, if it can be shown that  
a particle travels sufficiently far (about $ 180^\circ $) around
the black hole, then it follows that there can be causal contact between all points of the new universe at the time of the bounce. It is then probable that particles which fall into the black hole will interact frequently before reaching $r=0$, establish thermal equilibrium and thus homogenize.
Hence, when matter enters the new universe it will already be isotropic.

Here we will examine the geodesic equations for Schwarzschild geometry in order to follow 
the trajectory of a particle in orbit around a
black hole.  We will then search for orbits which exhibit the desired behavior.

The Schwarzschild metric is
\be\label{sm}
ds^2 = - \left(1 - \frac{2m}{r}\right) dt^2 
        + \frac{dr^2}{( 1 - \frac{2m}{r} )} + r^2 d\Omega^2
\,,
\ee
where $d\Omega^2$ is the metric on the unit sphere.  The equations of motion
are
\bea
\ddot t + \frac{2m}{r^2} \frac{1}{( 1 - \frac{2m}{r} )} \dot t \dot r & = & 0 \label{eom1}\\
\ddot r + \frac{m}{r^2} 
\left(1 - \frac{2m}{r}\right) \dot t^2 - r \sin^2\theta\left(1 - \frac{2m}{r}\right) \dot \phi^2 \nonumber \\
- r \left(1 - \frac{2m}{r}\right) \dot \theta^2 - \frac{m}{r^2} \frac{1}{( 1 - \frac{2m}{r} )} \dot r^2 & = & 0 \label{eom2} \\
\ddot \theta - \cos\theta\sin\theta \dot\phi^2 + \frac{2}{r} \dot\theta\dot r & = & 0 \label{eom3} \\
\ddot \phi + 2 \cot \theta \dot\phi\dot\theta + \frac{2}{r} \dot\phi\dot r & = & 0 \label{eom4}
\eea
where the dot indicates differentiation with respect to proper time $\tau$.

It is now possible to derive the equation for the orbit of a particle
moving in a Schwarzschild background, $\phi(r)$.  For simplicity we will take
the particle's trajectory to lie in the equatorial plane, $\theta = \pi/2$,
$\dot\theta = \ddot\theta = 0$.  The EOM (\ref{eom1})-(\ref{eom4}) simplify
and are easily solved to give
\be\label{julia}
\phi(r) = \int \frac{L\, dr}{r^2 \, \sqrt{E^2 - \left(1 - \frac{2m}{r}\right)\left(\frac{L^2}{r^2} + \k \right)}}
\,,
\ee
where $\k=1$ for matter and $\k = 0$ for photons.  Here we have defined constants
for the energy $E = (1-2m/r) \dot t$ and angular momentum $L=r^2 \dot \p$.\footnote{The
$E$ and $L$ are constants since the Lagrangian associated with the metric (\ref{sm}) is not explicitly
dependent on $t$ and $\p$.}  Upon making the substitution $u = 1/r$ equation (\ref{julia}) becomes
\be\label{pu}
\phi(u) = - \int \frac{du}{\sqrt{2mu^3 - u^2 + \frac{2m\k}{L^2}u-\frac{\k}{L^2} + \frac{E^2}{L^2}}}
\,.
\ee
For a rigorous treatment of the above integral see \cite{Chandra}.  The geometries of the
geodesics are determined by the roots of the function beneath the square root.  For
our purposes it is sufficient to make a few approximations.  We will break down the analysis
by studying three relevant regions around the black hole.  First we will consider
the behavior of orbits at large $r$.  Next, we consider orbits of particles
with initial position barely inside the
event horizon (i.e. with $r\approx 2m$ and $r < 2m$).  Finally we will consider orbits
deep inside the horizon close to the singularity.
\subsubsection{Small $u$ approximation}\label{secsu}

In the small $u$ (large $r$) regime the initial particle position is far from the 
black hole and the $u^3$ term
in (\ref{pu}) can be ignored.  Upon integration the equation of the orbit becomes
\be\label{smu}
\p(u) = \arcsin{\frac{2(\frac{m\k}{L^2} - u)}{\sqrt{-\Delta}}}
\,,
\ee
where $\Delta = -4(E^2 -\k + \frac{\k^2 m^2}{L^2})/L^2$.  For simplicity we will
consider photon trajectories and take $\k = 0$.  Hence,
\be\label{pup}
\p(u) = \arcsin{\left( \frac{-L u}{E}\right)}
\,.
\ee
This equation will admit orbits of the desired type.  We could have guessed this behavior since
in the large $r$ regime the orbits are appropriately described by Newtonian gravity.
Bound orbits which circle the black hole numerous times before falling into the
singularity will have ample time to interact with near horizon matter
and reach thermal equilibrium before
plunging to $r=0$.
\subsubsection{Near-horizon approximation}\label{secnh}

Consider orbits of photons starting \it just \rm inside the event horizon.
The least significant term in this regime is the $u^2$ term.  The integral (\ref{pu}),
with $\k = 0$ reduces to
\be\label{pun}
\phi(u) = - \int \frac{du}{\sqrt{2mu^3 + \frac{E^2}{L^2}}}
\,.
\ee
We start the photon at a distance of $r_i \approx 1.9 m$
and let it fall all the way to the singularity, $r_f = 0$.  We use the same values
for $E$, $m$ and $L$ as \cite{Chandra}.\footnote{Note that varying these values
does not seem to change the results significantly.}  
The results are that the photon travels through an angle of $\approx 114^\circ$ before
falling into the singularity.  Hence, some photons will be able to reach thermal equilibrium.  
Probably not enough, however, to solve the horizon problem.
As one would expect, the result turns out to be even less promising when we
start near the singularity.
\subsubsection{Large $u$ approximation}\label{seclu}

{}For very small values of $r$, far inside the horizon the dominant term in (\ref{pu})
is the $u^3$ term.  Keeping only this term and integrating we find
\be\label{lu}
\p(\Delta r) = \frac{2}{\sqrt{2m}} (\Delta r)^{1/2} 
\,,
\ee
where $\Delta r = {r_f}-{r_i}$. Here $r_i$ is the initial radial value of the particle
(with $r<2m$)
and $r_f$ is the final radius which we take to be extremely close to the singularity.

{}From (\ref{lu}), we deduce that orbits of this type are unable to ``bend" enough to traverse an angle $\p$ 
anywhere near the desired $180^\circ$.  For example, if we take $r_i = m$ (near the
largest $r$ where our approximation is still valid) and let
this particle fall to the singularity at $r_f = 0$ we find a maximum angle
of $\p \approx 80^\circ$.  It appears that in
order for the black hole mechanism described above to solve the horizon problem
we will need to rely on matter reaching equilibrium before it crosses the horizon.  Such
geodesics are discussed in section \ref{secsu}.
  
Hence we have successfully addressed the horizon problem. Firstly, the causality aspect of the horizon problem disappears because there is causal contact
between all points in space at the beginning of the de Sitter phase. In addition, matter contained in a homogeneous parent universe will necessarily produce a homogeneous daughter universe via 
the black hole mechanism discussed above.  Even if the parent universe is not homogeneous,
the matter falling into the black hole may have ample time to reach equilibrium
by following geodesics of the type described in 
section \ref{secsu}.  Matter which has not come into causal contact
by the time it has reached the horizon is less likely to achieve equilibrium 
before entering the de Sitter region (see \ref{seclu} and  \ref{secnh}).  Thus, it appears that one cannot solve the horizon problem by considering interactions
which occur solely inside the horizon.

\subsection{The Flatness Problem}

Another problem of the SBB is the flatness problem. Observations indicate
that a sizeable fraction of the critical matter density is contained in
galaxy clusters and therefore
$\Omega$, the parameter defined to be the ratio of the density of matter in the
universe measured today $\rho_0$ to the critical density $\rho_c$ needed to give a spatially flat universe, is of the order of $\Omega = 1$. Assuming an initial spectrum of adiabatic scale-invariant density fluctuations, the recent
observations of the cosmic microwave background (CMB) anisotropies 
\cite{Boomerang,Maxima} indicate that $\Omega$ is very close to 1, a value which
in standard cosmology is an unstable fixed point in an expanding universe.  

If the universe were not nearly flat, it is very likely that we would
not exist.  If $\Omega$ is larger than $1$ the universe would have collapsed
long ago, perhaps even on the order of the Planck time, $10^{-43}$ seconds.  
If $\Omega$ is less than $1$ the universe would have expanded too fast
to allow for structures such as galaxies to form.
As flatness is an unusual condition among the class of standard models
one might expect cosmology to reveal a mechanism which requires flatness (such as inflation).

It is possible to reformulate the flatness problem in terms of the parameter $k$ in
the Friedmann-Robertson-Walker line element of relativistic cosmology,
\be\label{frw}
ds^2 = - dt^2 + R^2(t) \left(\frac{dr^2}{1-kr^2} + r^2 (d\theta^2 + \sin^2{\theta} d\phi^2)\right)
\,.
\ee
Here $k$ gives the geometry of 3-spaces of constant curvature and can be
$-1, \, 0, \, 1$.  $R(t)$ is the scale factor.  If $k = -1$ or $0$, the universe is open and if $k=1$ the universe
is closed and will collapse.  It is possible to calculate the energy density
of the universe at the present instant $\rho_0$ in terms of $k$ by using the Einstein equations
and the metric (\ref{frw}),
\be\label{dens}
\rho_0 = \frac{3}{8\pi G} \left(\frac{k}{R_0^2} + H_0^2 \right)
\,.
\ee
The flatness problem now translates into explaining why $k=0$, since 
$\rho_c = \frac{3H_0^2}{8\pi G}$.  

The argument we use to solve the flatness problem (i.e. explain why $k=0$)
is very simple.  In order to smoothly connect the black hole exterior to the 
de Sitter interior the spatial topology
of both regions must be the same.  Therefore, if the parent black hole
has one topology, so must the daughter universe,
generated from the mechanism discussed above. In our model we are cutting out the innermost region of a black hole manifold and matching it to half of de
Sitter space. Whereas the space-time topology of de Sitter space is fixed, the topology of the spatial sections depends on the space-time slicing selected by the physics (see e.g. the textbook \cite{HE} for a detailed discussion of this 
issue). We will now argue that in our construction we must match the inside of 
a black hole manifold to the flat spatial sections of de Sitter space (given 
by the coordinates ${\hat t}, {\hat x}, {\hat y}, {\hat z}$ in \cite{HE}).  

\begin{figure}[htbp]
\includegraphics[angle=-90,width=3.5in]{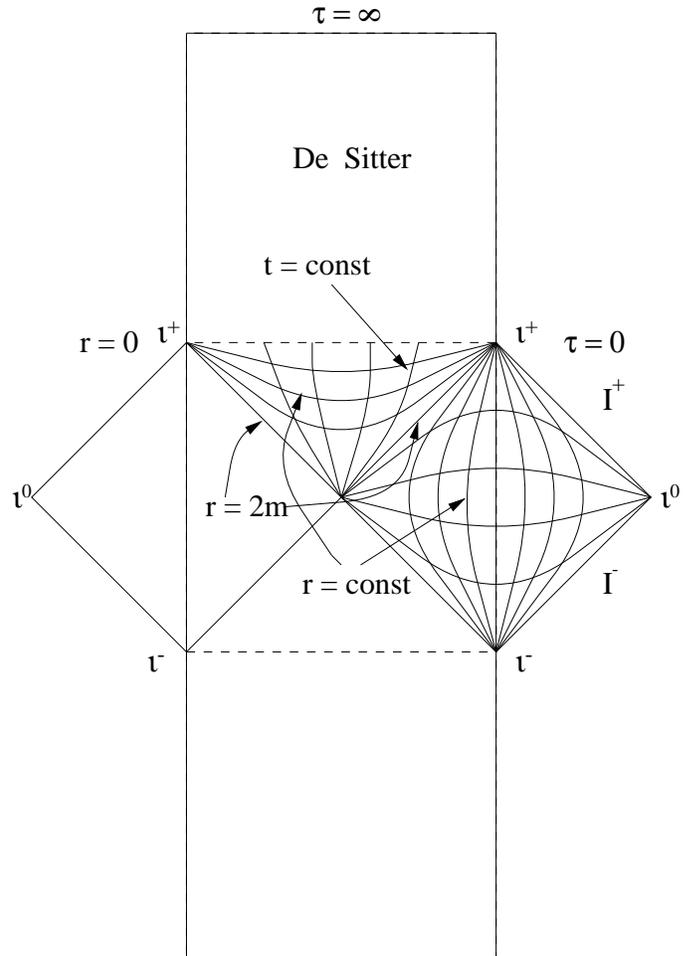}
\caption{
The space-time diagram of a nonsingular Schwarzschild black hole with de
Sitter interior.  The length of the dotted line gives the spatial
size of the universe.  The location of this line is $r=0$ in the
Schwarzschild coordinates and corresponds to the initial time surface
$\tau = 0$ of the de Sitter regime.  Note that the roles of $r$ and
$t$ switch when crossing the horizon.}
\label{pen2}
\end{figure}

The full space-time 
diagram of the model is sketched in Fig. (\ref{pen2}). 
In this scenario, the spatial size of the daughter
universe is given by the length of the dotted line.  If the daughter
universe is infinite in spatial extent the parameter $k$ must be
either $-1$ or $0$.  This length is easily calculated 
using the Schwarzschild metric.  Note, that as one crosses the event horizon
space and time change roles.  With the same inevitability that an observer
outside of the black hole must move forward in time, grow old and die an observer inside the horizon can only move forward in ``space" eventually reaching the center
of the black hole.  It is therefore sensical to make the variable substitution
$r \rightarrow - t$ in the Schwarzschild metric, and the interior metric (with $\theta = 0$ and $\phi = const$)
becomes
\be\label{inmetric}
ds^2 = \left( 1 + \frac{2m}{t} \right) dr^2 - \frac{dt^2}{\left( 1 + \frac{2m}{t} \right)}
\,.
\ee
Note that the horizon is now located at $t = - 2m$.  To calculate the size of the
universe we may integrate along a line of constant $t$ near $t=0$ over
all space
(see Fig. (\ref{pen2})):
\be\label{l}
s = \int_{- \infty}^\infty \sqrt{\left( 1 + \frac{2m}{t} \right)} \, dr
\,,
\ee
which is clearly infinite.  

It is possible to match the above result with a calculation performed from the de Sitter perspective.
For a locally de Sitter universe, the scale factor for the metric $(\ref{frw})$ grows like, 
\begin{eqnarray}
& & \sinh{H\tau} \qquad (k = -1) \nonumber \\
R  & \propto & \cosh{H\tau} \qquad (k = 1) \\
& & \exp{H\tau} \qquad (k = 0) \nonumber
\end{eqnarray}
Here we have introduced the variable $\tau$ to represent the
time coordinate of the de Sitter Universe.  Thus, the size of the spatial sections of the de Sitter universe is given
by integrating $ds$ along a surface of constant $\tau$ (see Fig. (\ref{pen2})):
\be\label{lds}
s = \int_{- \infty}^\infty \frac{R(\tau) \, dr}{\sqrt{1 - k r^2}}
\,.
\ee
Note that the integral (\ref{l}) tends to infinity linearly.  The only value for $k$ in 
(\ref{lds}) which causes the integral to linearly approach infinity is the $k = 0$, flat solution. Therefore, in order to match the spatial topology of the de Sitter universe to that of the black hole we must be matching to the spatially flat slices of de Sitter space, i.e. have $k=0$. 
 
\subsection{The Structure Formation Problem}

It is generally assumed that structures in the universe were formed by gravitational
instabilities from some initial perturbations.  Explaining the origin of these perturbations is one of the main challenges for modern cosmology, a challenge
which inflationary cosmology quite successfully meets: quantum fluctuations produced during the de Sitter phase on sub-Hubble scales are the seeds for the observed perturbations (see e.g. \cite{MFB} for a comprehensive review).

In the scenario discussed above in which our universe emerges as the inside of a black hole via a short de Sitter bounce, a quantitative theory of the
origin and evolution of fluctuations remains to be worked out. Here, we mention some ideas. 

First, it is possible that cosmic string defects generated during a matter phase transition in the new universe provide the seeds for structure (see e.g. \cite{VS,HK,RHBrev} for comprehensive reviews). At the moment, it appears that such a mechanism has problems explaining the observed \cite{Boomerang,Maxima} narrow first ``Doppler" peak in the spectrum of CMB anisotropies. 

It may be possible that Hawking radiation of the parent black hole could 
provide a solution to the structure formation
problem in the new universe. The radiation appears as quantum fluctuations in the de Sitter space.
If the wavelength of a fluctuation becomes greater than the particle horizon of
the de Sitter space $(\l >> c/H)$ causality forces the fluctuation to become frozen as 
a classical amplitude that can seed structure formation.
 
\subsection{The Black Hole Information Loss Problem}

Yet another problem which is naturally solved in this scenario is the black hole information
loss problem.  Consider a large star of 
mass $M$ which is initially described by a pure quantum state $\ket{\psi}$ having zero entropy.
Now imagine this star undergoes gravitational collapse and forms a black hole.  Before the collapse
the density matrix of the star is given by $\rho = \ket{\psi} \bra{\psi}$.
After collapse, the black hole begins to decay via thermal Hawking radiation and
is described by a mixed state, with corresponding density matrix
\be\label{eq:den}
\rho = \sum_{n} p_n \ket{\psi_n} \bra{\psi_n} \,.
\ee
Here the $p_n$ are the probabilities of having different initial states.
Since the entropy of the Hawking radiation is of order $M^2/m_{pl}^2$ where $m_{pl}$ 
is the Planck mass, the final state is missing information as compared to the initial
state.  If the black hole decays via radiation until it completely disappears it will
take any information about the matter which has fallen into it with it.  This
leads to a fundamental contradiction with the laws of quantum mechanics which state
that a pure state cannot evolve into a mixed state.  In the scenario depicted 
above all pure states which fall into the black hole
emerge in the new universe as pure states and the information loss problem
is avoided.

\vspace{0.4cm}
\section{Discussion and Conclusions}

In this paper we have studied consequences of cosmological scenarios in 
which our universe is born from a black hole resting in a parent universe \cite{FMM} - \cite{DKH}.
We have discussed the ways in which some of the problems of the SBB
model may be solved given such a model.  In particular, we have provided
a solution to the horizon problem by examining geodesics of matter falling
into a black hole and showing that it is possible to bring this matter into causal contact
before it emerges in the new universe.  

A possible solution to the flatness problem was discussed.  The size of the daughter
universe was calculated from both the Schwarzschild and the de Sitter perspectives.  The
Schwarzschild calculation predicts the universe to be infinite.  The integral
representing the length approaches infinity linearly.  In order to recover the same result in the de Sitter frame, we must be matching the interior of the black hole to the spatially flat sections of de Sitter space. This is the
only way to make the length approach infinity linearly.  Hence, it appears
that our model singles out the observed, flat FRW universe via a
topological argument.

We showed that this scenario does not suffer from 
the black hole information loss problem since pure states evolve 
to pure states and information is transferred
from the parent universe to the black hole interior universe.  Finally,
a relation between structure formation and Hawking radiation was suggested.

There are a number of interesting problems which will be left for future research.
One is to construct a higher derivative 
theory of gravity which 
will provide a realization of the above mentioned model.  The new gravitational action 
should admit a solution which
resembles the Schwarzschild black hole at large distances but with the
singularity replaced by a de Sitter universe.  The construction will
be similar to that in \cite{EB}.  It will be interesting to consider 
applications of the AdS/CFT correspondence to such models.  We would also like
to compare thermodynamical quantities
of the exterior black hole to those of the de Sitter universe inside.  
The effects of black hole evaporation on the universe may prove interesting.
\vspace{0.4cm}

\centerline{\bf Acknowledgments}
\vspace{0.4cm}
DE is grateful to A. Gosh, L. Smolin and G. Veneziano for useful discussions.
DE was supported in part by the 
U.S. Department of Education under the GAANN program. RB was
supported in part by the U.S. Department of Energy under
Contract DE-FG02-91ER40688, TASK A.

\end{document}